\newcommand{\Kshort}{\ensuremath{K^0_S}}

\newcommand{\MeV}{\ensuremath{\textrm{MeV}}}

\newcommand{\GeV}{\ensuremath{\textrm{GeV}}}

\documentclass[
    ,final            
  ]
  {aipproc}

\layoutstyle{8x11single}

\usepackage[latin1]{inputenc}
\usepackage{ae,aecompl} 
\usepackage{subfig}

\begin{document}

\title{Study of kaonic final states in $\pi^-p$ at $190\,\textrm{GeV}$}


\author{Tobias Schlüter for the COMPASS collaboration}{
  address={Department für Physik, Ludwig-Maximilians-Universität, Am
    Coulombwall 1, 85748 Garching, Germany}
}

\begin{abstract}
  We discuss the status of analyses of data recorded in the 2008 and
  2009 runs of the COMPASS experiment at CERN with sepcific focus on
  final states with $\Kshort\Kshort\pi^-$ and $K^+K^-\pi^-$ produced
  in $\pi^-(190\,\GeV)p$ scattering.  The interest in such final
  states is motivated by a summary of some of the relevant literature.
  We also show first results from the analysis of diffractively
  produced $K\bar K\pi$ states.  Two prominent three-body structures,
  one around $1.8\,\GeV$, the other at $2.2\,\GeV$ decaying via known
  $K\bar K$ and $K\pi$ states are seen.
\end{abstract}

\maketitle


\section{Introduction}
\label{sec:introduction}

In light-meson spectroscopy final states including two kaons are
interesting for several reasons.  First, final state kaons mean higher
thresholds, allowing cleaner spectroscopy of the heavier states.
Second, no states with exotic quantum numbers have so far been
observed in kaonic final states.  Third, strange mesons are an
interesting research topic by themselves.  We shall outline our
specific interests below.

The physics goals of the COMPASS~\cite{Abbon:2007pq} hadron runs are
spectroscopy of mesons from central and diffractive production.
Specific aims are the confirmation or rejection of hybrid or glueball
states.  The literature on both is vast.  At the time of this writing,
the SPIRES database finds 782 papers with ``glueball'' in their title.
Out of these, 6 also have ``evidence'' in their title.  But only two
of these papers have been cited more than 5 times: one is a lattice
calculation~\cite{Sexton:1995kd}, the other~\cite{Amsler:1995tu} was
followed by a paper by the same authors~\cite{Amsler:1995td} phrasing
their result as a question.  Likewise a database search finds 200
papers on mesons with ``hybrid'' in their title.  Yet, there's not a
single one whose title contains ``evidence'' or ``confirm.''  In other
words, clarification and further experimental input are direly needed.

The COMPASS collaboration recorded diffractive scattering data on a
liquid hydrogen target with both negative and positive hadron beams at
$190\,\GeV$ during its 2008 and 2009 beam times.  The forward-flying
particles produced in the central or diffractive interaction were
measured and identified by means of a two-stage spectrometer which
features highly efficient track reconstruction from approximately
$1\,\GeV$ upwards combined with particle identification in the
wide-angle spectrometer and both electromagnetic and hadronic
calorimetry in both stages of the spectrometer.  Together, these
components yield nearly $4\pi$ coverage of the neutral and charged
particles of the forward flying system.  Further, in order to identify
events where the target proton remained intact, a recoil proton
detector was used which detected the slow proton emitted at large
angles.  At the same time a veto system ensured the absence of other
slow particles between the recoil proton detector and the spectrometer
acceptance, thought mostly due to inelastic excitations of the target.

In the COMPASS experiment kaons can be identified in essentially the
following ways without recourse to physical processes:
\begin{itemize}
\item charged kaons can be recognized by means of a RICH detector,
  which measures the angle of Cherenkov radiation emitted by particles
  passing through its gas ($\mathrm{C}_4 \mathrm{F}_{10}$) volume.
  This angle is a function of the velocity $\beta$ of the particles.
  Since track reconstruction in magnetic spectrometers measures the
  momentum $p$ of the particles, their mass $m$ can be recovered via
  $p/m = \beta/(1 - \beta^2)^{1/2}$.  In our current analysis we use
  this to identify charged kaons between $10\,\GeV$ and $30\,\GeV$.
  Figure~\ref{fig:rich} serves to illustrate the performance of the RICH
  detector.
\item short-lived neutral kaons identify themselves through their
  displaced decay $V_0$ vertex which gives the right mass if a pion
  hypothesis is made for its outgoing oppositely charged particles.
  Given the known~\cite{Amsler:2008zzb} mean lifetime $\tau =
  0.9\times 10^{-10}\,\textrm{s}$ and the neutral kaon mass
  $m(\Kshort) = 497\,\MeV$, this displacement can take the order of a
  few meters at COMPASS energies.  The most important background to
  the \Kshort{}s are $\Lambda \to p\pi$ decays.  Their contribution
  was verified to be negligible.  Reconstructed \Kshort-masses are
  shown in fig.~\ref{fig:mKshort}.  Note that the exclusive sample
  used in the analysis has very little background.
\item finally, the beam carries a fraction of kaons ($O(5\%)$).  These
  are excluded by means of specifically designed Cherenkov
  detectors~\cite{Bovet:1975bx} which were placed in the beamline.
\end{itemize}

\begin{ltxfigure}
\subfloat[Cherenkov angle versus particle momentum.  Three bands
    appear.  These correspond to different mass particles, from left
    to right: pions, kaons and (anti-)protons.  The peak in the upper
    left corner is due to $\delta$-electrons.]{
      \includegraphics[width=.42\textwidth]{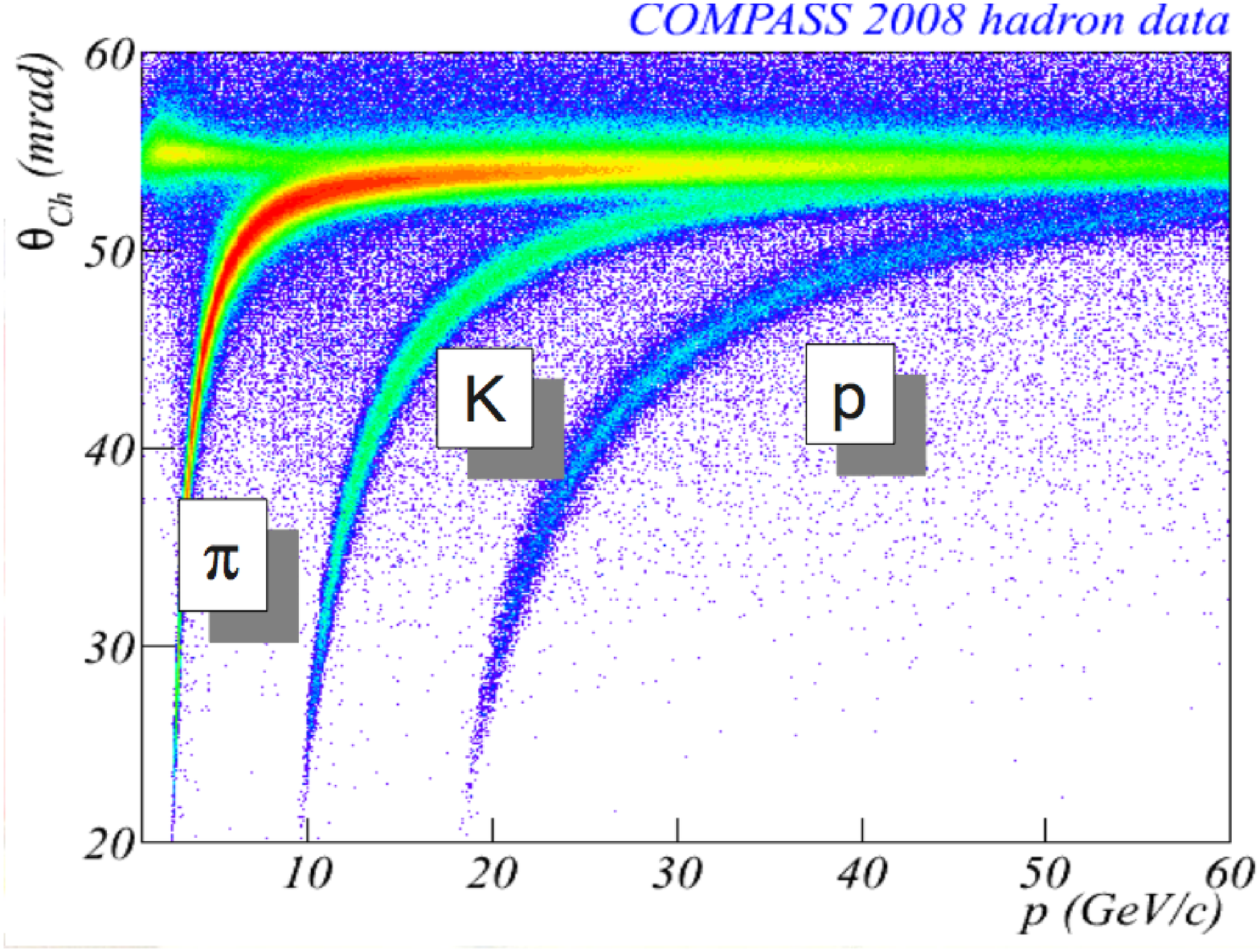}
      \label{fig:rich}}
\hfill
  \subfloat[Distribution of reconstructed \Kshort{} masses before and
    after (filled histogram) exclusivity cuts (explained further down
    in text).]{
  \label{fig:mKshort}
  \includegraphics[width=.48\textwidth]{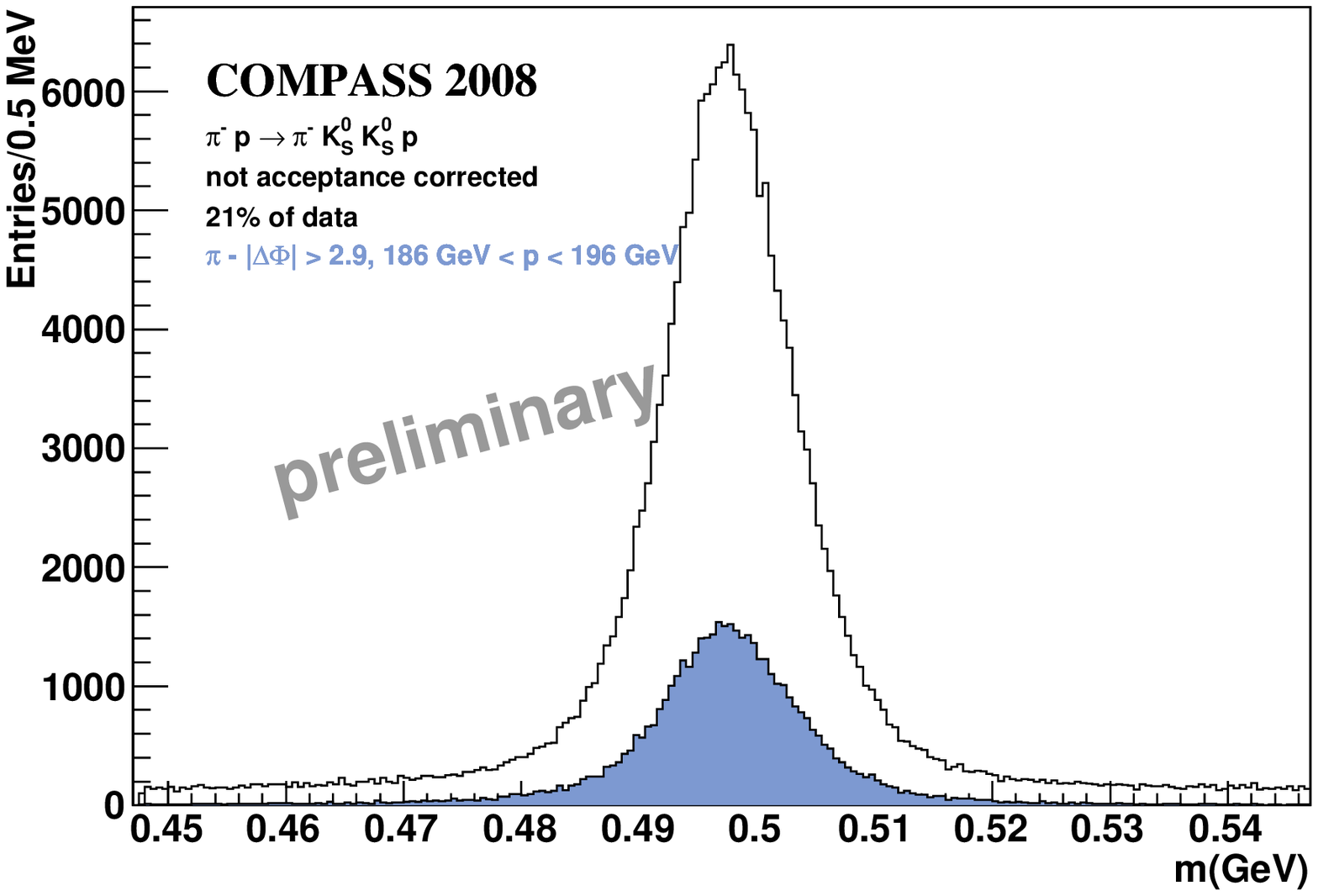}}
\caption{Kaon identification}
\end{ltxfigure}

In this paper we discuss the current state of our analyses of the
exclusive $K^+K^-\pi^-$ and $\Kshort\Kshort\pi^-$ systems produced in
diffractive scattering of a $\pi^-$ on a proton.  The centrally
produced $K\bar K$ subsystem in the same channel was discussed
previously~\cite{Schluter:2009,Schott:2009} where we also discussed
the different quantum numbers accessible to $K^+K^-$ and
$\Kshort\Kshort$.

\section{Physics expectation}
\label{sec:physics-background}

The $K \bar K \pi$ system, as produced in $\pi^- p$ collisions, can
exist with exotic quantum numbers $1^{-+}$.  Preferential $K^*(\to K
\pi) K$ decay is expected from models of quartet and hybrid
states~\cite{Chung:2002fz,Close:1994hc}.  So far no kaonic exotics
emerged from various experimentally studied production
processes~\cite{Klempt:2007cp}.  Since their existence is inevitable
in flavor multiplet schemes, they can be considered touchstones of
quartet models.  COMPASS produced $K\bar K\pi^-$ with high statistics,
well suited for partial wave analysis.  In addition to exotics, states
or dublets with $q \bar q$ quantum numbers like the enigmatic
$E/\iota(1405)$ or $\pi(1800)$ will possibly make an appearance in
these data.

The question whether the $\pi(1800)$ is a radial excitation of the
pion or a hybrid state or a combination of both, perhaps even two
different nearby states, has been raised.  Theoretical calculations of
branching ratios in the different scenarios~\cite{Barnes:1996ff},
quoted in table~\ref{tab:branchings}, have provided a reference frame
for experiments to distinguish between the various possibilities.
Results in this direction were given by VES~\cite{Amelin:1997uu} and
E852~\cite{Eugenio:2008zz}.

\begin{table}
\begin{tabular}[c]{lccccccr}
  \hline
  & $\rho\pi$ & $\rho\omega$ & $\rho(1465)\pi$ & $f_0(1300)\pi$ & $f_2\pi$ &
  $K^*K$ & Total \\
  \hline
  $\pi_{3S}(1800)$ & 30 & 74 & 56 &   6 & 29 & 36 & 231 \\
  $\pi_H(1800)$    & 30 & 0  & 30 & 170 & 6 & 5 & $\approx 240$ \\
  \hline
\end{tabular}
\caption{Predicted branching of $\pi(1800)$ for different model
  assumptions~\cite{Barnes:1996ff}.  Note the much
  suppressed $K^*K$ branching for a strictly hybrid $\pi(1800)$
  compared to the case of a radial excitation.}
\label{tab:branchings}
\end{table}

There are several contested states around $2.1\,\GeV$ that are left
out of the PDG summary tables.  One of special interest to us is the
$f_2(2150)$ which was found to decay to $a_2(1320)\pi$ by the Crystal
Barrel collaboration~\cite{Adomeit:1996nr}.  Following the
systematization given in~\cite{Anisovich:2000kxa,Anisovich:2008} the
$f_2(2150)$ should be an $s\bar s$ state and a radial excitation of
the $f_2(1525)$ which decays predominantly to $K\bar K$.  Therefore,
we expect the $f_2(2150)$ to also decay via $K^*\bar K$ and $\bar K^*
K$.

\section{Data selection}
\label{sec:data-selection}

The data shown was selected from two weeks of the COMPASS 2008 hadron
run.  Details on the selection can be found
in~\cite{Schluter:2009,Schott:2009}.  The quality of the
reconstruction and selection is illustrated with the exclusivity plots
from the neutral kaon data, shown in fig.~\ref{fig:exclusivity}.  One
sees that momentum conservation by itself ensures a very clean
selection owing to the purity of the COMPASS trigger system and the
efficient track reconstruction.  The coplanarity angle referenced is
defined as the angle between the plane spanned by the beam direction
and the momentum vector of the $K\bar K\pi$ system on the one
hand, and the plane spanned by the beam direction and the recoiling
proton on the other.  These planes should coincide by momentum
conservation.  Due to the segmented architecture of the recoil proton
detector a precision of $\approx 0.15\,\textrm{rad}$ is expected.
This expectation is nicely matched by the data.

\begin{ltxfigure}
  \subfloat[Total reconstructed momentum of the $\Kshort\Kshort\pi^-$ data
    set.]{
      \includegraphics[width=.48\textwidth]{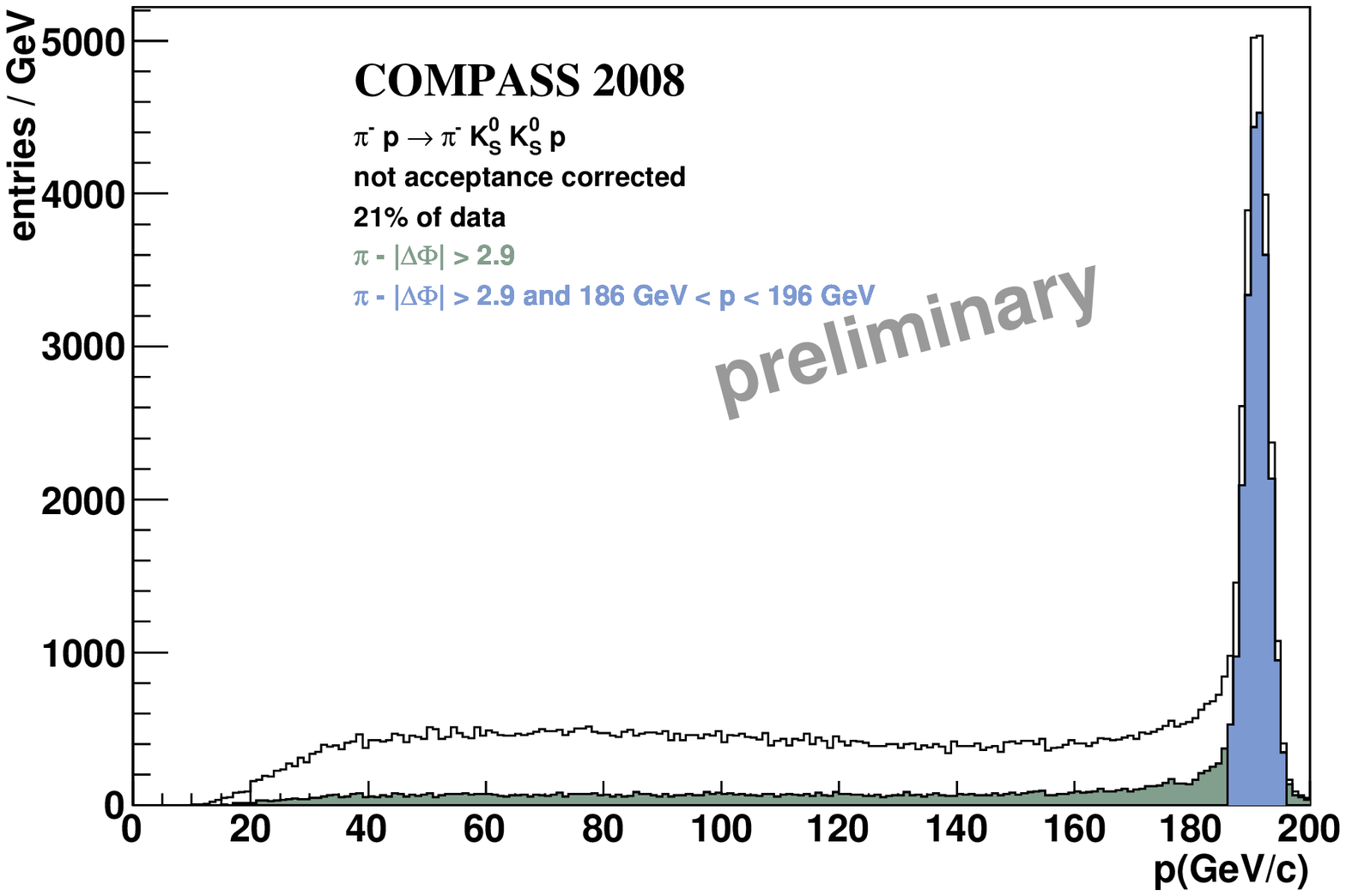}
      \label{fig:pTot}
    }
  \quad
  \subfloat[Coplanarity angle difference of the $\Kshort\Kshort\pi^-$ data
    set.]{
      \includegraphics[width=.48\textwidth]{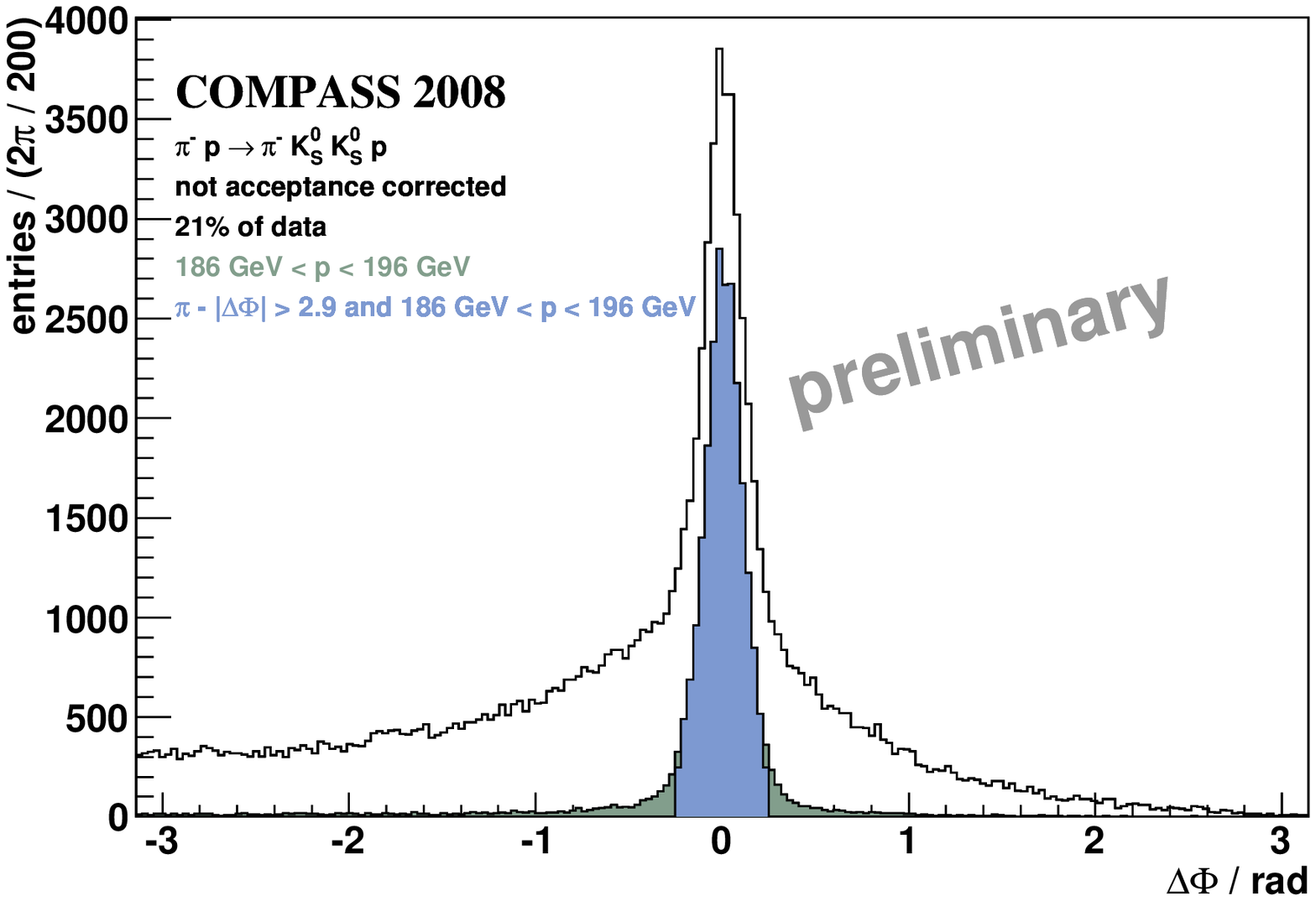}
      \label{fig:deltaphi}
    }
    \caption{Exclusivity of the $\Kshort\Kshort\pi^-$ data sample.
      After the topological selection, two further cuts were applied:
      momentum within $5\,\GeV$ of the nominal $191\,\GeV$ and a
      coplanarity angle $\Delta\Phi$ (defined in text) such that $\pi
      - |\Delta\Phi| > 2.9$.  Both plots show the total sample, and
      the effect of each exclusivity cut on the other variable.  The
      final set with both cuts is also indicated.}
  \label{fig:exclusivity}
\end{ltxfigure}

For the partial wave analysis, we kinematically fitted the $\Kshort$
vertices to a mass hypothesis and included the so-obtained neutral
tracks into the fit of the primary vertex in a way similar to the
techniques given in~\cite{Forden:1985xa,Luchsinger:1992ia}.  The plots
shown are without these fits but the conclusions are not affected.

One remark concerns the use of the RICH detector in the charged kaon
set.  For the time being, we imposed a cut at $30\,\GeV$ above which
we found that kaon identification became unreliable.  This infers a
significant cut in low 3-body masses where momentum conservation
dictates that all final state particles have momenta well above
$30\,\GeV$.  Imposing the same cut on the $\Kshort$ data makes most
differences between the pictures shown below disappear.

\section{Status of analysis}
\label{sec:first-impr-gain}

\begin{ltxfigure}
  \subfloat[Rapidity distribution of the $\Kshort\Kshort\pi^-$.  There
  are two \Kshort{} entries per event.]{
    \includegraphics[width=.48\textwidth]{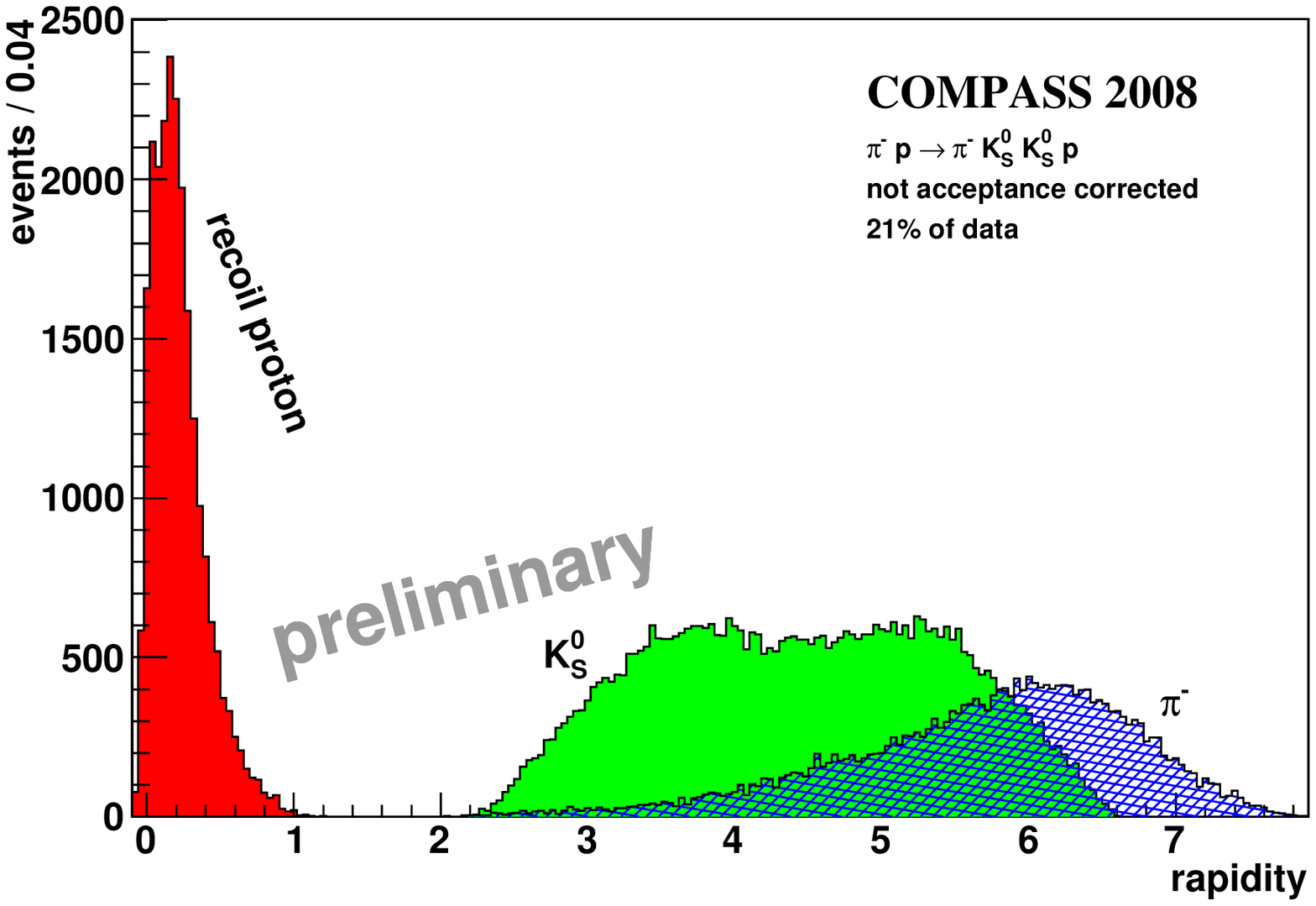}
    \label{fig:rap}
  }
  \quad
  \subfloat[$\Kshort\Kshort$ mass distribution as function of the $\pi$
  momentum.  Note the division into different production regimes.]{
    \includegraphics[width=.48\textwidth]{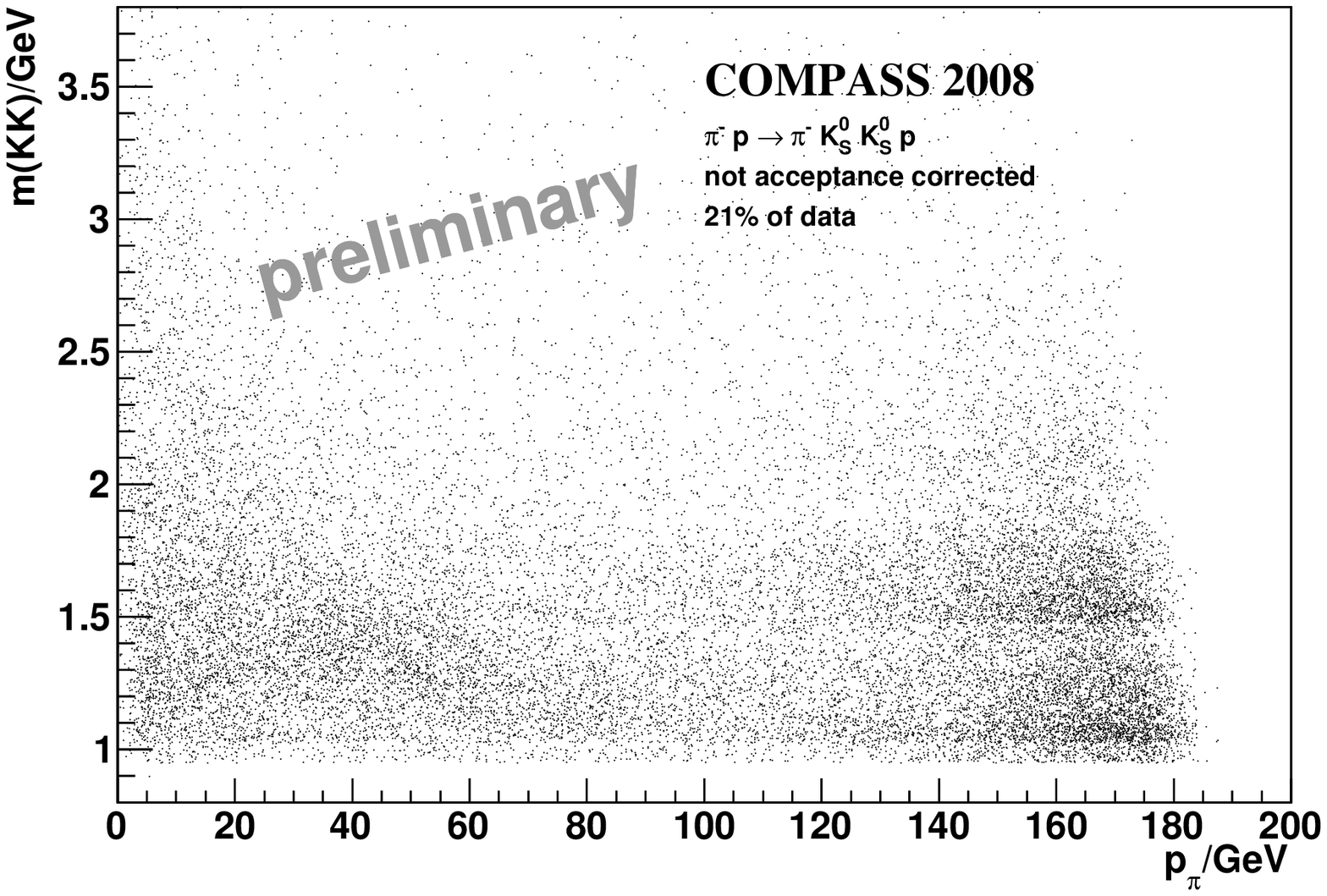}
    \label{fig:mKK_vs_ppi}
  }
  \caption{Illustration of different production regimes in the case of
    the neutral kaon data.  (Not acceptance corrected.)}
  \label{fig:dynamics}
\end{ltxfigure}

\begin{ltxfigure}
  \subfloat[$m(K^+K^-\pi^-)$ against $m(K^+\pi^-)$]{
    \includegraphics[width=.48\textwidth]{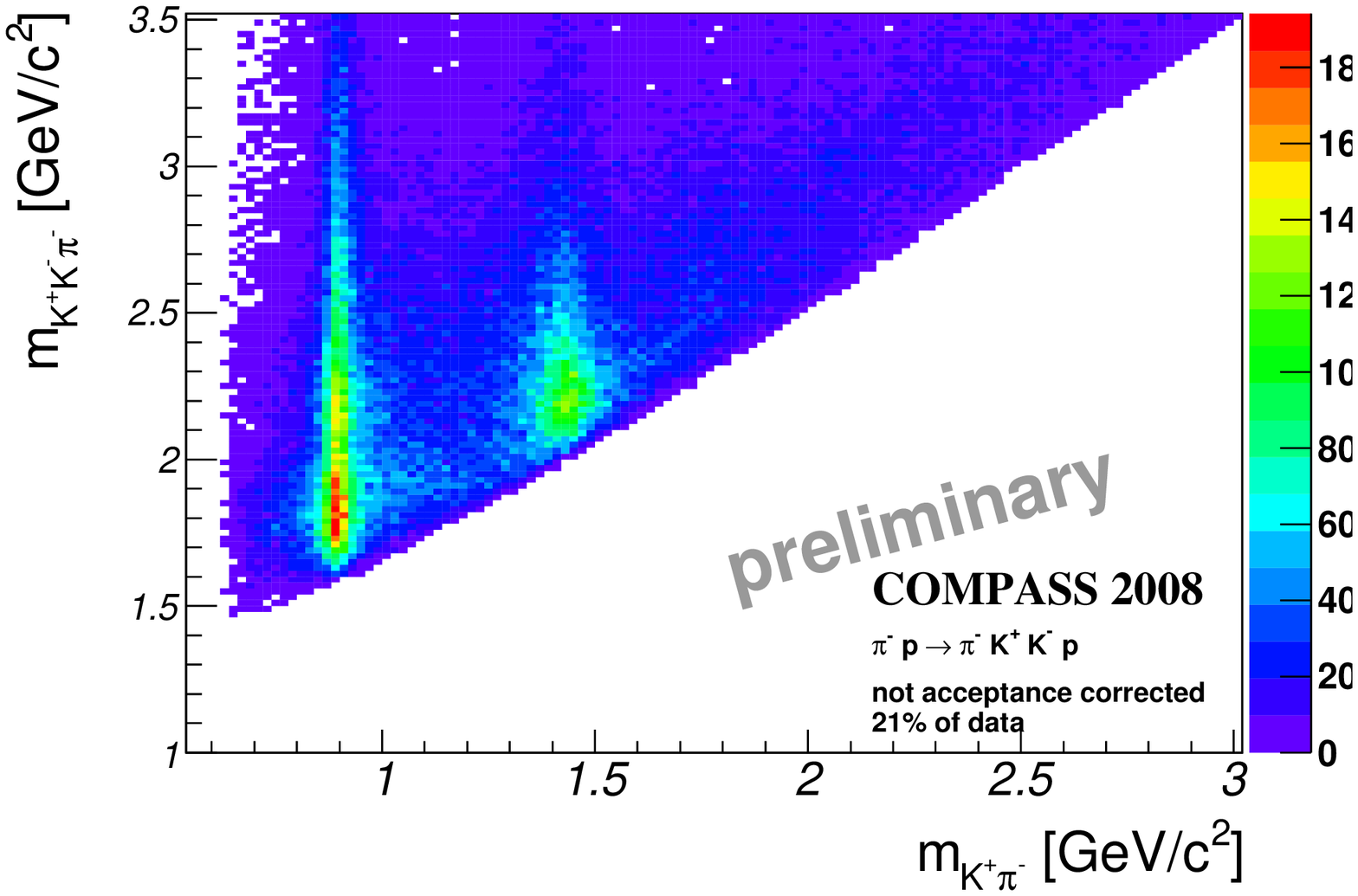}
    \label{fig:tobi_KplPi}
  }
  \subfloat[$m(\Kshort\Kshort\pi^-)$ against $m(\Kshort\pi^-)$ (two
  entries per event)]{
    \includegraphics[width=.48\textwidth]{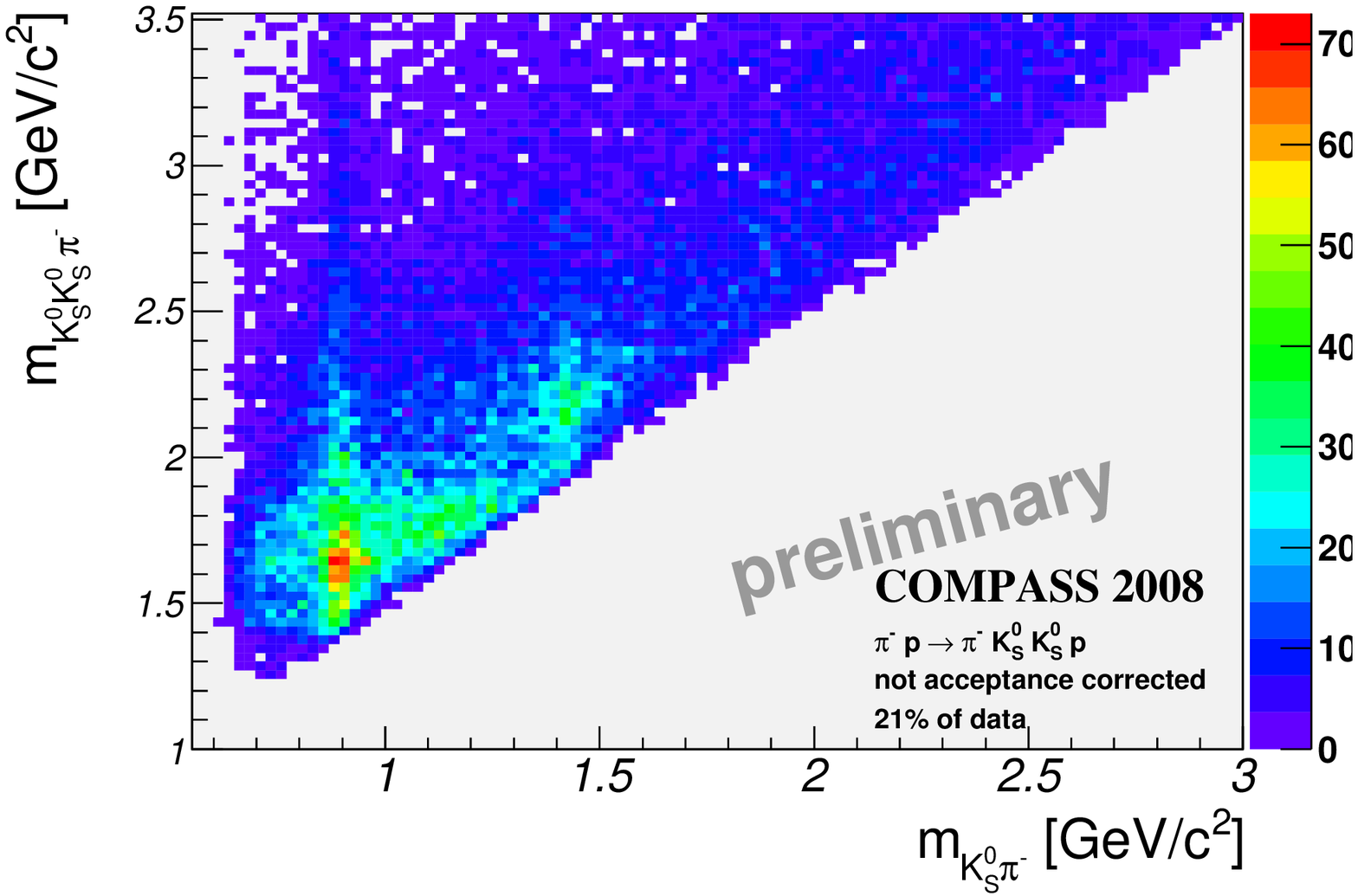}
    \label{fig:tobi_KsPi}
  }
  \subfloat[$m(K^+K^-\pi^-)$ against $m(K^+K^-)$]{
    \includegraphics[width=.48\textwidth]{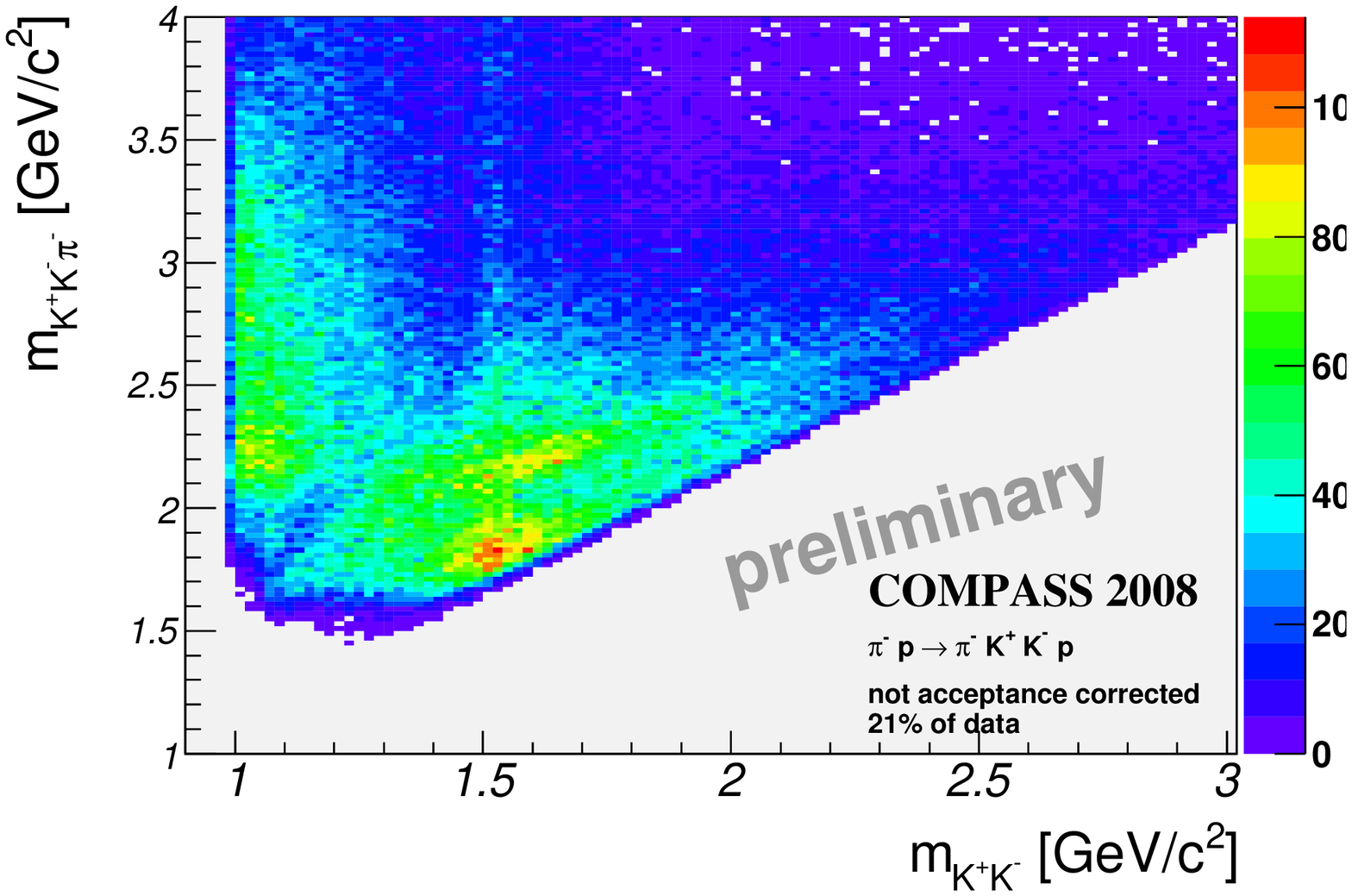}
    \label{fig:tobi_KplKm}
  }
  \subfloat[$m(\Kshort\Kshort\pi^-)$ against $m(\Kshort\Kshort)$]{
    \includegraphics[width=.48\textwidth]{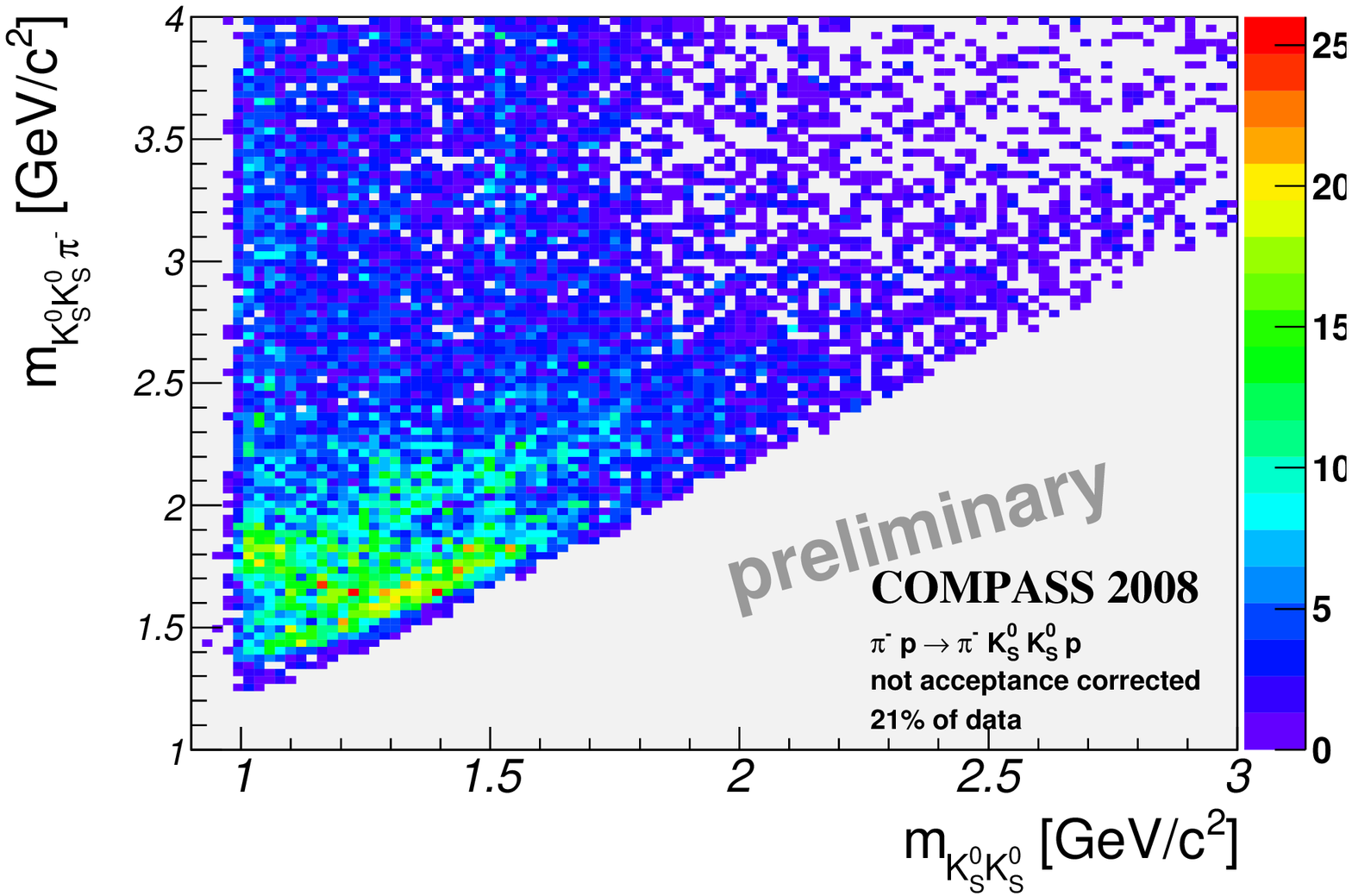}
    \label{fig:tobi_KsKs}
  }
  \caption{Two-body invariant masses against three-body invariant
    masses for the various possible combinations.  All plots show
    three-body states around $1.8\,\GeV$ and $2.1\,\GeV$ selectively
    decaying through the various visible two-body states.  Comparing
    figs. (a) and (b) shows the dramatic loss in the lower left corner
    of (a) due to the loss of high-momentum kaons explained in the
    text.  (Not acceptance corrected.)}
  \label{fig:tobi}
\end{ltxfigure}

The presence of different production regimes is apparent in the
neutral kaon data.  Figure~\ref{fig:dynamics} illustrates this in two
ways.  The distribution of rapidities (fig.~\ref{fig:rap}) shows large
overlap between the pion and kaon rapidities, but also a significant
fraction where the pion rapidity exceeds that of the kaons.
Diffractive production is characterized by the absence of a rapidity
gap between the decay products of the diffractively excited resonance.
For the partial wave analysis we selected a subset without an apparent
rapidity gap.  The division into different production processes is
corroborated by fig.~\ref{fig:mKK_vs_ppi} which shows the $K\bar K$
mass distribution over the momentum of the pion.  For high pion
momenta, the $K\bar K$ mass spectrum develops detailed structure
whereas for lower pion momenta the $K\bar K$ mass spectrum appears to
be dominated by the available phase space.

Figure~\ref{fig:tobi} gives a detailed view of the mass spectra in the
overall sample and how they relate.  The first striking feature
concerns the $K\pi$ spectra (top row).  There are very strong bands
corresponding to the $K^*(892)$ and the excited kaons (which we shall
indiscriminately refer to as $K(1430)$ in what follows) around
$1430\,\MeV$.  There are two predominant structures in terms of the
three-body mass: one at approximately $1800\,\MeV$, one around
$2.2\,\GeV$.  The lower two plots, which allow the same kind of
quantitative analysis for $K\bar K$ intermediate states, show two
strong diagonal bands which are reflections of the $K^*(892)$ and
$K(1430)$.  Besides these, one sees the $f(980)$ / $a(980)$ near the
threshold and a well defined band near $m(K\bar K)=1500\,\MeV$ which
probably corresponds to the $f_2(1525)$.  Again the clustering near
$m(K\bar K\pi) = 1800\,\MeV$ and $= 2.2\,\GeV$ is apparent.  The
difference in structure between the charged and the neutral case is
significantly reduced if one reproduces the low-momentum cut needed
for the charged kaons in the neutral case.

A preliminary partial wave analysis of the structure near $1800\,\MeV$
was done.  The resulting partial waves turned out to be compatible
with a significant $\pi(1800)$ contribution.  The structure at
$2.2\,\GeV$ appears consistent with spin 2.  No further or definite
conclusion can be drawn at our current state of PWA.

\section{Outlook}
\label{sec:conclusion}

We have shown promising results from our preliminary analyses of
diffractively produced $K\bar K\pi^-$ states.  Refining our selection
and incorporating the complete data set of the 2008 and 2009 COMPASS
runs will put us in a position which will allow partial wave analyses
encompassing large parts of the known and unknown light meson
spectrum.  We have shown that our partial wave software is in a state
that allows first preliminary conclusions but a significant amount of
works remains to be done, both in understanding our acceptance and in
implementing the various intermediate states.  Yet, the future looks
bright.

\begin{theacknowledgments}
  This research was supported by the DFG cluster of excellence 'Origin
  and Structure of the Universe' (www.universe-cluster.de).  The
  author would like to express his gratitude to D. Ryabchikov for his
  lessons in PWA software given during several Sunday afternoons.
\end{theacknowledgments}



\bibliographystyle{aipproc}   

\bibliography{partialwaves,general}

\end{document}